\PassOptionsToPackage{table}{xcolor}
\documentclass[sigconf]{acmart}
\copyrightyear{2026}
\acmYear{2026}
\setcopyright{cc}
\setcctype{by}
\acmConference[RecSysChallenge '26]{RecSys Challenge 2026}{October 02, 2026}{Minneapolis, MN, USA}
\acmBooktitle{RecSys Challenge 2026 (RecSysChallenge '26), October 02, 2026, Minneapolis, MN, USA}
\acmDOI{10.1145/3842413.3842414}
\acmISBN{979-8-4007-2863-1/2026/10}

\newcommand{\Gr}[1]{{\color{gray}#1}}

\begin{document}

\title[Overview and Analysis of the RecSys Challenge 2026: Conversational Music Recommendation]{Overview and Analysis of the RecSys Challenge 2026: Conversational Music Recommendation}

\author{Seungheon Doh}
\orcid{0000-0002-8448-9928}
\affiliation{\institution{Sony Group Corporation}\country{Japan}}
\email{seungheon.a.doh@sony.com}

\author{Sergio Oramas}
\orcid{0000-0002-8028-2890}
\affiliation{\institution{SiriusXM}\country{United States}}
\email{sergio.oramas@siriusxm.com}

\author{Bruno Sguerra}
\orcid{0000-0003-1158-9095}
\affiliation{\institution{Deezer Research}\country{France}}
\email{bmassonisguerra@deezer.com}

\author{Abhinav Bohra}
\orcid{0009-0008-8377-0604}
\affiliation{\institution{Amazon}\country{United States}}
\email{bohraabhinav8@gmail.com}

\author{Claudio Pomo}
\orcid{0000-0001-5206-3909}
\affiliation{\institution{Politecnico di Bari}\country{Italy}}
\email{claudio.pomo@poliba.it}

\author{Francesco Barile}
\orcid{0000-0003-4083-8222}
\affiliation{\institution{Maastricht University}\country{Netherlands}}
\email{f.barile@maastrichtuniversity.nl}

\begin{abstract}
The RecSys Challenge 2026 studies conversational music recommendation as a joint item recommendation and response generation problem: given a multi-turn dialogue, systems must retrieve relevant tracks from a large catalog and produce a grounded natural-language response. This paper presents the challenge task, dataset, evaluation protocol, and official results. Beyond the leaderboard, we analyze the 16 accepted systems through a common retrieve--rerank--generate framework and examine how recommendation performance varies across users, requests, and dialogue contexts. Strong systems commonly combine heterogeneous candidate sources and preserve source-specific evidence for learned reranking. Across the system papers and our organizer-side analysis, robust design also means 1) grounding cold-start retrieval in multi-turn conversation and item signals, 2) using intent detectors, and 3) modeling the full multi-turn context rather than the current query alone. We further identify limitations of the benchmark and evaluation protocol, including single-ground-truth relevance and teacher-forced evaluation of synthetic dialogues. Together, these findings provide practical guidance for future conversational recommender systems and shared evaluation efforts.
\end{abstract}




\maketitle

\begin{minipage}[t]{\columnwidth}
\centering
\includegraphics[width=0.95\columnwidth]{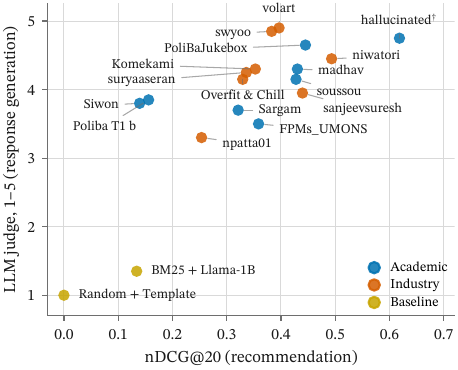}
\captionof{figure}{Challenge at a glance: recommendation quality (nDCG@20) versus response quality (LLM-as-a-judge, 1--5). $\dagger$ The evaluation-data-exposed submission.}
\label{fig:teaser}
\raggedright

\section{Introduction}
\vspace{3mm}
\end{minipage}

The RecSys Challenge 2026 focuses on conversational music recommendation~\cite{epure2025music, oramas2024talking, palumbo2025text2tracks, doh2025talkplay, doh2025talkplaytools}. Conversational recommender systems (CRS) let users express, refine, and receive recommendations through natural, multi-turn dialogue, replacing the static, one-shot lists of traditional recommenders~\citep{goker2000adaptive, christakopoulou2016towards, zhang2018towards}. This is equally critical for music listening: through dialogue, listeners can express their musical preferences, refine them turn by turn, and come to understand the music itself, rather than specifying a fixed preference upfront. A request can identify an exact title, describe an acoustic texture or cover image, ask for a mood or activity, or react implicitly to tracks played earlier, so a useful system must retrieve precisely from a large catalog, update its belief about intent over turns, and explain its choices without inventing evidence. Realizing this requires coupling natural language understanding with high-precision recommendation: at every turn the system must perform two tightly coupled functions, (i) \textit{music recommendation}, ranking relevant tracks from a large catalog given the dialogue history and user context, and (ii) \textit{response generation}, producing a natural-language reply that justifies the recommendation and sustains the conversation.

This paper first documents the 2026 task, dataset, evaluation protocol, and official results (Sections~\ref{sec:task}--\ref{sec:evaluation}). We then go beyond the leaderboard in three ways. Section~\ref{sec:approaches} surveys the 16 accepted systems and their retrieval, reranking, and response-generation components. Section~\ref{sec:protocol} revisits every submission by dialogue depth, warm versus cold condition, goal specificity, and goal topic, using organizer-side resources that were not released to participants (the unmasked source conversations, their goal annotations, and the hidden target tracks). Sections~\ref{sec:principles} and~\ref{sec:limitations} distill the systems into reusable design principles and audit the benchmark itself, including an evaluation-data exposure in the winning submission and limitations of the dataset and protocol.

Our analysis leads to four main conclusions. (1) Strong systems combine candidates from complementary retrievers and learn from source-specific ranks, scores, and presence together with cross-source agreement. (2) Cold-start pipelines should rely by default on conversation and item signals that remain available without a user ID or profile. (3) Rather than attempting to recover a complete hidden goal taxonomy, systems should use high-precision, abstaining intent detectors to trigger specific ranking actions. (4) Recommendation intent is distributed across turns, so systems should model the full dialogue rather than focus only on the current query. Together, these findings offer an evidence-based blueprint for designing future conversational music recommendation systems.

\section{The Challenge Task}
\label{sec:task}

\vspace{2mm}\noindent\textbf{Problem Formulation.}
Let $\mathcal{D} = \{d_1, \dots, d_N\}$ be the track catalog, where each track $d_j$ carries multimodal content: metadata (title, artist, album, release date), semantic tags, lyrics, audio, and album image. Each user $u$ is associated with a profile $p_u$ of demographic attributes (age group, gender, country). A conversation between the user and the system unfolds over $T$ turns as $s = \{(q_i, m_i, r_i)\}_{i=1}^{T}$, where at each turn $i$ the user issues a natural-language query $q_i$, and the system responds with a ranked list of tracks $m_i \subset \mathcal{D}$ together with a natural-language response $r_i$. The task is to model the system at each turn. Concretely, at turn $t$ a system $f$ is given the user profile $p_u$, the conversation history $s_{t-1} = \{(q_i, m_i, r_i)\}_{i=1}^{t-1}$, and the current query $q_t$, and must produce both a ranked recommendation list and a response $(m_t, r_t) = f(p_u, s_{t-1}, q_t)$. The two outputs define complementary subtasks.

\vspace{2mm} \noindent \textbf{Task 1. Music Recommendation.}
The recommendation output $m_t$ is a ranked list of the top 20 tracks from $\mathcal{D}$, which must capture both the explicit request and the implicit preferences expressed across the conversation history $s_{t-1}$, the current query $q_t$, and musical taste of user $u$.

\vspace{2mm} \noindent \textbf{Task 2. Response Generation.}
The response output $r_t$ is a natural-language reply that justifies the recommended tracks $m_t$ and maintains a coherent conversational flow with the user, conditioned on $(p_u, s_{t-1}, q_t)$.

\vspace{2mm}\noindent\textbf{Participation.}
The challenge ran on Codabench~\footnote{https://www.codabench.org/competitions/15786} from April 10 to June 30, 2026, with separate academic and industry tracks and two evaluation phases, interim (Blind-A) and final (Blind-B). It attracted 336 registered teams and 9,554 submissions, of which 40 teams reached the Blind-B leaderboard; 20 teams submitted system papers, of which 16 were accepted after review. The organizers also released two reference baselines: a random recommender with template responses, and BM25 retrieval over the dialogue paired with a Llama-1B response generator.

\section{Dataset}
\label{sec:dataset}

We source conversation sessions from the \textit{TalkPlayData-Challenge} dataset~\cite{choi2025talkplaydata}, a synthetic corpus constructed via a multimodal agentic pipeline grounded in the LFM-2b dataset~\cite{schedl2022lfm2b}, in which a Listener LLM and a Recommender LLM engage in a goal-conditioned dialogue, yielding sessions of up to eight turns grounded in real user demographics, listening histories, and track metadata. The data is split into three parts: participants train and validate on a public release\footnote{\url{https://huggingface.co/datasets/talkpl-ai/TalkPlayData-Challenge-Dataset}} of 15{,}199 training and 1{,}000 development conversations, and are then ranked on two held-out blind sets of 80 sessions each, \textit{Blind-A}\footnote{\url{https://huggingface.co/datasets/talkpl-ai/TalkPlayData-Challenge-Blind-A}} and \textit{Blind-B}\footnote{\url{https://huggingface.co/datasets/talkpl-ai/TalkPlayData-Challenge-Blind-B}}. Ground-truth responses and target tracks are never disclosed for the blind sets, which are drawn from distinct conversations to measure generalization rather than overfitting: Blind-A drives the live leaderboard, while Blind-B, released near the deadline, determines the final ranking. Blind-B additionally masks the user identifier and profile in 40 of its 80 sessions; for analysis we call the profile-visible half \emph{warm} and the masked half \emph{cold}, and Section~\ref{sec:limitations} explains why this is an operational missing-profile condition rather than a fully disjoint-user cold-start test.

\begin{table}[t]
\centering
\caption{Statistics of the TalkPlayData-Challenge splits. \#Tracks is the number of distinct tracks. \#Turns is the mean turns per conversation; $|q|$ and $|r|$ are mean query and response lengths in words.}
\label{tab:stats}
\resizebox{\columnwidth}{!}{%
\begin{tabular}{lrrrrrr}
\toprule
Split & \#Conv. & \#Users & \#Tracks & \#Turns & $|q|$ & $|r|$ \\
\midrule
Training Set & 15{,}199 & 8{,}591 & 43{,}597 & 8.0 & 44.7 & 36.6 \\
Dev Set & 1{,}000 & 500 & 6{,}761 & 8.0 & 40.8 & 44.6 \\
Blind-A Set & 80 & 58 & 208 & 3.6 & 39.0 & 33.5 \\
Blind-B Set & 80 & 54 & 274 & 4.5 & 42.1 & 35.2 \\
\bottomrule
\end{tabular}%
}
\end{table}

\begin{table*}[t]
\centering
\small
\caption{Official Blind-B results for all teams with an accepted system paper. Recommendation diversity is catalog coverage, lexical diversity is Distinct-2, and the LLM-as-a-judge score is on its original 1--5 scale. $\dagger$ The evaluation-data-exposed submission.}
\label{tab:leaderboard}
\begin{tabular}{rllrrrrr}
\toprule
 & & & & \multicolumn{2}{c}{Recommendation System} & \multicolumn{2}{c}{Response Generation} \\
\cmidrule(lr){5-6}\cmidrule(lr){7-8}
Rank & Team & Track & Final Score & nDCG@20 & C. Diversity & LLM Judge & L. Diversity \\
\midrule
\rowcolor{gray!10} \Gr{1} & \Gr{hallucinated~\cite{alkhetiar2026talk}$^{\dagger}$} & \Gr{Academic} & \Gr{0.689} & \Gr{0.618} & \Gr{0.027} & \Gr{4.750} & \Gr{0.957} \\
2 & volart~\cite{volgin2026frozen} & Industry & 0.587 & 0.397 & 0.032 & 4.900 & 0.927 \\
3 & niwatori~\cite{wakatsuki2026practical} & Industry & 0.586 & 0.493 & 0.031 & 4.450 & 0.774 \\
4 & swyoo~\cite{yoo2026twoviews} & Industry & 0.578 & 0.383 & 0.030 & 4.850 & 0.952 \\
5 & PoliBaJukebox~\cite{lops2026distribution} & Academic & 0.576 & 0.445 & 0.030 & 4.650 & 0.765 \\
9 & madhav~\cite{patil2026routing} & Academic & 0.548 & 0.430 & 0.031 & 4.300 & 0.817 \\
11 & soussou~\cite{soussou2026multisource} & Academic & 0.530 & 0.428 & 0.031 & 4.150 & 0.771 \\
12 & sanjeevsuresh~\cite{suresh2026label} & Industry & 0.528 & 0.440 & 0.032 & 3.950 & 0.833 \\
16 & Komekami~\cite{terai2026auditing} & Industry & 0.500 & 0.353 & 0.028 & 4.300 & 0.731 \\
17 & suryaaseran~\cite{seran2026dialogue} & Industry & 0.492 & 0.336 & 0.029 & 4.250 & 0.773 \\
20 & Overfit \& Chill~\cite{sundrani2026minimaestro} & Industry & 0.477 & 0.330 & 0.031 & 4.150 & 0.733 \\
24 & FPMs\_UMONS~\cite{manderlier2026picking} & Academic & 0.451 & 0.359 & 0.031 & 3.500 & 0.810 \\
25 & Sargam~\cite{rai2026dialogue} & Academic & 0.448 & 0.321 & 0.012 & 3.700 & 0.840 \\
29 & npatta01~\cite{pattaniyil2026state} & Industry & 0.381 & 0.254 & 0.031 & 3.300 & 0.786 \\
31 & Poliba T1 b~\cite{valentini2026graph} & Academic & 0.366 & 0.156 & 0.031 & 3.850 & 0.710 \\
34 & Siwon~\cite{lee2026empirical} & Academic & 0.354 & 0.140 & 0.031 & 3.800 & 0.713 \\
\midrule
-- & BM25 + Llama-3.2-1B-Instruct & Baseline & 0.162 & 0.134 & 0.020 & 1.350 & 0.664 \\
-- & Random + Template Response & Baseline & 0.067 & 0.000 & 0.033 & 1.000 & 0.632 \\
\bottomrule
\end{tabular}
\end{table*}

To let participants focus on modeling rather than feature extraction, we provide four supporting resources over the shared catalog of 47{,}071 tracks and 8{,}772 users. \textit{Track Metadata}\footnote{\url{https://huggingface.co/datasets/talkpl-ai/TalkPlayData-Challenge-Track-Metadata}} and \textit{User Metadata}\footnote{\url{https://huggingface.co/datasets/talkpl-ai/TalkPlayData-Challenge-User-Metadata}} provide the official track catalog (title, artist, album, tags, etc.) and the demographic attributes of the user profile $p_u$. We further release pre-extracted \textit{Track Embeddings}\footnote{\url{https://huggingface.co/datasets/talkpl-ai/TalkPlayData-Challenge-Track-Embeddings}} spanning six modalities: audio (CLAP~\cite{wu2023large}), album image (SigLIP-2~\cite{tschannen2025siglip}), text-based tags, lyrics, and metadata (Qwen3-Embedding 0.6B~\cite{zhang2025qwen3}), and a collaborative-filtering embedding (BPR~\cite{rendle2009bpr}). Matching \textit{User Embeddings}\footnote{\url{https://huggingface.co/datasets/talkpl-ai/TalkPlayData-Challenge-User-Embeddings}} are also provided for the collaborative-filtering side.

Each conversation goal has two axes~\cite{choi2025talkplaydata}. The topic code covers audio (A, e.g.\ ``discover songs with immersive soundscapes''), lyrics (B), visual--musical connections (C), contextual or situational use (D), interactive refinement (E), metadata-rich exploration (F), mood and emotion (G, e.g.\ ``I need something to cheer me up''), artist/discography (H), culture/geography (I), social/popularity context (J), and temporal/era discovery (K, e.g.\ ``music from the 80s please''). Specificity is a two-letter code: the first letter is query specificity and the second target specificity. LL (e.g.\ ``play some chill music'') and HL (e.g.\ ``find bebop jazz with saxophone, 1950s--60s'') admit many relevant tracks; LH (e.g.\ ``what was the popular song from a recent musical movie?'') describes a vague clue with one or few intended targets; and HH (e.g.\ ``Wonderwall by Oasis'') is a precise lookup.

\section{Evaluation}
\label{sec:evaluation}

The official score combines what the system recommends and how it describes the recommendation: $\mathrm{Score}=0.50\,\mathrm{nDCG@20}+0.10\,\mathrm{Catalog\ Diversity}+0.10\,\mathrm{Lexical\ Diversity}+0.30\,\mathrm{LLM Judge}_{\mathrm{norm}}$.

For each evaluated session-turn $e$, only one track is labeled relevant. If it appears at rank $k\leq20$, its nDCG contribution is $1/\log_2(k+1)$, and otherwise zero. The official aggregate first averages within each target-turn index and then across turn indices. Blind-B deliberately contains ten targets at each of turns 1--8, so its official macro average equals the ordinary 80-session mean.

Catalog diversity is the fraction of catalog tracks that appear at least once across all submitted lists. Lexical diversity is corpus-level Distinct-2~\cite{li2016diversity}. On a fixed subset of 10 Blind-B sessions, the LLM judge~\cite{zheng2023judging,doh2026llmjudge} scores personalization and explanation quality from 1 to 5; the two dimension means are averaged and normalized as $(\mathrm{Judge}-1)/4$ for the score above. The final evaluator used Gemini-3.1-Flash-Lite~\cite{gemini31}.

\vspace{2mm}\noindent\textbf{Official results.}
Table~\ref{tab:leaderboard} groups the component metrics by subtask. The winner, hallucinated, leads ranking by a wide margin (0.618 nDCG@20 versus 0.493 for the next system). Although the team did not use Blind-B's hidden ground-truth answers, its paper discloses that the Blind-B dataset itself was used for training: fusion weights were tuned and the final reranker trained on its conversation turns. We flag this evaluation-data exposure by shading throughout this paper (Section~\ref{sec:limitations}). The two subtasks are only loosely coupled: the best judge scores belong to volart (4.90) and swyoo (4.85), whose rankings are mid-pack, and rank-2 volart trails rank-3 niwatori by 0.10 nDCG while beating it by 0.45 judge points. Every accepted system beats both organizer baselines on ranking, although the margin is small for the lowest-ranked entries. Catalog diversity is nearly constant (0.03) because 80 sessions $\times$ 20 slots cap it at 0.034 of the 47,071-track catalog, so it does not discriminate between systems~\cite{volgin2026frozen}.

\section{Submitted Approaches}
\label{sec:approaches}

\begin{table*}[t]
\centering
\setlength{\extrarowheight}{0.8mm}
\footnotesize
\vspace{-2mm}
\caption{Component summary of the 16 accepted systems, ordered by final rank: retrieval (candidate generation), reranking/fusion, and response generation. Multi-component pipelines are condensed to their load-bearing pieces. $\dagger$ The evaluation-data-exposed submission.}
\label{tab:methods}
\begin{tabular}{>{\raggedright\arraybackslash}p{2.2cm}>{\raggedright\arraybackslash}p{4.9cm}>{\raggedright\arraybackslash}p{4.7cm}>{\raggedright\arraybackslash}p{4.2cm}}
\toprule
Team & Retriever (Candidate Generation) & Reranker & Response generator \\
\midrule
\rowcolor{gray!10} \Gr{hallucinated~\cite{alkhetiar2026talk}$^{\dagger}$} & \Gr{BM25, dense, hybrid two-tower, ItemKNN, six multimodal sequential generators, all CVaR-tuned} & \Gr{Weighted RRF; XGBoost LTR ensemble; continuity/query heuristics} & \Gr{Gemma-4-26B-A4B-it, summarize--generate--compress--diversify} \\
volart~\cite{volgin2026frozen} & BM25, OpenAI text-embedding-3-large, LLM track-description, co-occurrence & RRF pool; LambdaMART with missing-field-safe features & Best-of-3 Gemini-3.1-Flash-Lite; independent gpt-4o-mini critic edit \\
niwatori~\cite{wakatsuki2026practical} & 14 lexical, dense, entity, history, transition, co-occurrence sources & LightGBM LambdaRank with source features & Qwen3.6-27B grounded on top-3; diversity selection \\
swyoo~\cite{yoo2026twoviews} & BM25 + frozen Qwen3-Embedding-0.6B hybrid pool; QLoRA-tuned Qwen3-Embedding-8B conversational pool & Top-100 union; LightGBM with cross-pool RRF and request-regime features & GLM-5.2 propose--assign--select with deterministic claim validation \\
PoliBaJukebox~\cite{lops2026distribution} & 10 lexical, collaborative, CLAP acoustic, organizer Qwen3 embeddings and fine-tuned BGE-base/BGE-large sources & Weighted RRF; LightGBM LambdaRank + CatBoost YetiRank; density-ratio weighting & Two-pass fact-grounded generation (Gemini-3.1-Pro-Preview draft, Gemini-3.5-Flash critic) \\
madhav~\cite{patil2026routing} & Fine-tuned MPNet dense retrieval with continuity boosts & Rank-zone fence (popularity/co-occurrence); LightGBM tail reranker & Gemini-2.5-Pro narrates only the clean top-3 head \\
soussou~\cite{soussou2026multisource} & Broad, feedback-aware, and fielded BM25; fine-tuned BGE-base-en-v1.5; structural; visual & LightGBM LambdaRank with source features & Qwen3-14B multi-candidate; heuristic selection; constrained rewrite \\
sanjeevsuresh~\cite{suresh2026label} & BM25, dense text (metadata/lyrics/tags), query-expansion and album-artwork arms & XGBoost LambdaRank (49 features); exact-request detector gates a specialist trained with request-satisfying targets & LLM responder (model N/A) \\
Komekami~\cite{terai2026auditing} & 8 lexical anchor + 12 continuity/neighbor/entity/transition sources & Anchor LightGBM; LightGBM LambdaRank on 73 rank-only features & Claude Opus 4.8 on top-1 metadata; ranking frozen \\
suryaaseran~\cite{seran2026dialogue} & BM25, LoRA-tuned multilingual-e5-base and frozen Qwen3-Embedding-0.6B, history-neighbor, artist expansion, CF-BPR, co-occurrence, SASRec & LightGBM LambdaMART with 67 features & Decoupled responder (model N/A) \\
Overfit \& Chill~\cite{sundrani2026minimaestro} & Qwen3-8B query planner; 10 text, anchor, identity, training-mined channels & LightGBM LambdaRank with source-aware features & Same Qwen3-8B with three state-routed templates \\
FPMs\_UMONS~\cite{manderlier2026picking} & RRF over BM25, fine-tuned Qwen3-Embedding-4B, item-CF, artist expansion & Qwen3-8B listwise scoring head over top-200 & Best-of-20 Gemma-3n-E4B selected by a Gemma-4-E2B judge. \\
Sargam~\cite{rai2026dialogue} & Artist/album resolver; BM25; multi-seed CF/audio/visual and user-CF neighbors & Weighted RRF with exact-target override & Constrained Gemini response on the top-1 track (version N/A) \\
npatta01~\cite{pattaniyil2026state} & 11 BM25, dense-text, CLAP text-to-audio, anchor-centroid, user-CF, lookup branches & LambdaMART over branch, bi-encoder cosine and typed-state features & Qwen3-30B-A3B-Instruct-2507 conditioned on typed state \\
Poliba T1 b~\cite{valentini2026graph} & Heterogeneous user--session--track GNN; BM25; CLAP audio; Qwen3 lyric embeddings & Query-dependent neural router + weighted RRF; LLM rerank + reflective replacement & Reasoning draft + persona/style rewrite (model N/A) \\
Siwon~\cite{lee2026empirical} & BM25; frozen Qwen3-Embedding-4B; listening-history centroid & RRF & Frozen Qwen3-32B (thinking) rationale, frozen Phi-4 summarizer \\
\bottomrule
\end{tabular}
\vspace{1mm}\parbox{0.96\textwidth}{\footnotesize $^{\dagger}$The evaluation-data-exposed submission (Section~\ref{sec:limitations}).}
\vspace{-2mm}
\end{table*}

The 16 accepted papers span systems and diagnostics, but converge on a modular retrieve--rerank--generate cascade. Table~\ref{tab:methods} summarizes each team's retrieval, reranking, and response-generation components. Among the non-exposed systems, runner-up volart pairs a low-cost four-lane retrieval pool with LambdaMART and critic-guided response editing, arguing that ranking is information-limited once the candidate neighborhood is found, and posts the highest LLM-judge score of any submission~\cite{volgin2026frozen}. Third-place niwatori uses 14 retrieval sources and shows that retaining each source's rank, score, and presence is more effective than collapsing the union too early, giving it the strongest or joint-strongest ranking results among non-exposed systems on several goal topics (Section~\ref{sec:protocol})~\cite{wakatsuki2026practical}. swyoo's two-view retrieval and claim-validated generation pipeline places fourth with the second-highest judge score of any submission~\cite{yoo2026twoviews}, and PoliBaJukebox combines ten retrieval sources with adversarial-validation-based correction for dev-to-blind deployment shift~\cite{lops2026distribution}. The top-ranked submission, hallucinated, combines broad lexical, dense, two-tower, nearest-neighbor, and sequential candidate generators, each tuned under a CVaR objective, with weighted RRF fusion and an XGBoost ranker~\cite{alkhetiar2026talk}; however, its paper discloses that fusion weights and reranker were tuned and trained on Blind-B conversation turns, so we treat its numbers as descriptive only and set its results in gray throughout (Section~\ref{sec:limitations}).

\subsection{Retriever and Reranker}

Nearly every strong submission uses heterogeneous candidate sources. Common components are fielded BM25 and exact entity matching, dense query--track retrieval, artist/album expansion, item or session co-occurrence, and collaborative signals. Sargam adds multi-seed CF, audio, and visual neighbors and a high-specificity catalog resolver~\cite{rai2026dialogue}; FPMs\_UMONS combines BM25, dense retrieval, item-CF, and artist expansion~\cite{manderlier2026picking}; Poliba T1 b's graph-aware entry joins a heterogeneous user--session--track encoder with sparse, audio, and lyric retrievers under a learned router~\cite{valentini2026graph}.

The dominant fusion pattern is a high-recall union followed by a tree-based learning-to-rank model. niwatori, volart, PoliBaJukebox, swyoo, suryaaseran, Komekami, soussou, Overfit \& Chill, npatta01, and hallucinated all preserve source-specific evidence for LightGBM or XGBoost reranking~\cite{wakatsuki2026practical,volgin2026frozen,lops2026distribution,yoo2026twoviews,seran2026dialogue,terai2026auditing,soussou2026multisource,sundrani2026minimaestro,pattaniyil2026state,alkhetiar2026talk}. FPMs\_UMONS instead trains an LLM scoring head and shows that selecting the top item and ordering ranks 2--20 are distinct skills~\cite{manderlier2026picking}. Weighted RRF remains a strong training-free fallback and candidate-pool constructor~\cite{rai2026dialogue,lee2026empirical}.
Two regime-aware ideas are particularly reusable. PoliBaJukebox detects dev-to-blind covariate shift by adversarial validation and trains its rerankers with tempered density-ratio importance weights, a training-time correction that leaves inference unchanged~\cite{lops2026distribution}. madhav's Compass fences noisier popularity/co-occurrence signals into ranks 4--20 so that the generated response describes only a clean head, and its analysis shows that gating continuity by predicted continuation-versus-discovery intent recovers the novel-artist slice~\cite{patil2026routing}. These methods adapt to deployment regime and intent rather than assuming one fusion rule is universally correct.

\subsection{Grounded Response Generation}

Most teams decouple ranking from prose: once top-20 tracks are fixed, the generator receives visible dialogue and verified metadata. swyoo's propose--assign--select pipeline groups candidate evidence before writing and validates title and attribution claims deterministically~\cite{yoo2026twoviews}. volart and FPMs\_UMONS generate several responses, then apply independent critics or deterministic filters for grounding and diversity~\cite{volgin2026frozen,manderlier2026picking}. soussou uses constrained rewrites only when a heuristic quality score improves~\cite{soussou2026multisource}. The common lesson is that the language model may describe catalog facts and conversational evidence but must not silently change the ranked decision.

\section{Recommendation Performance Analysis}
\label{sec:protocol}

\begin{figure}[t]
\centering
\includegraphics[width=\columnwidth]{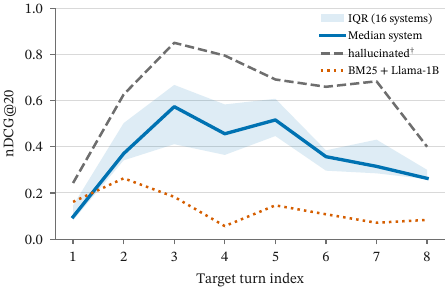}
\vspace{-5mm}\caption{Turn-macro nDCG@20 by target turn index over Blind-B. $\dagger$ The evaluation-data-exposed submission.}
\label{fig:turncurve}
\end{figure}

We join four organizer-side resources by $(\mathrm{session\_id},\mathrm{turn})$: public Blind-B inputs, unmasked source conversations with goal annotations, the hidden target file, and each submitted prediction. A session is cold-start exactly when the public Blind-B row masks its user ID/profile. For every system and slice, we compute nDCG@20 per session, average within each represented target turn, and then average across turns, matching the official macro convention. Warm/cold uncertainty uses 10,000 bootstrap resamples within each condition--turn cell. Goal-topic results are descriptive because topics contain only 2--12 sessions.

\subsection{Turn Depth}

Figure~\ref{fig:turncurve} suggests that longer dialogue contexts remain difficult for current systems: median nDCG rises from 0.10 at turn 1 to 0.57 at turn 3, but falls to 0.26 by turn 8. Turn 1 offers no played-track history, so continuation and co-occurrence sources are silent and systems depend entirely on the first request. The later decline matches Siwon's development finding that nDCG@20 falls monotonically with turn index and that request-focused query construction does not recover it~\cite{lee2026empirical}, and motivates the dual-view retrieval (a literal view of the latest request beside a full-dialogue view) used by several teams~\cite{soussou2026multisource,yoo2026twoviews}. The BM25 baseline stays low at every depth, which is consistent with history-aware sources contributing to the mid-conversation gains, although this comparison does not isolate their causal effect.

\begin{figure}[t]
\centering
\includegraphics[width=\columnwidth]{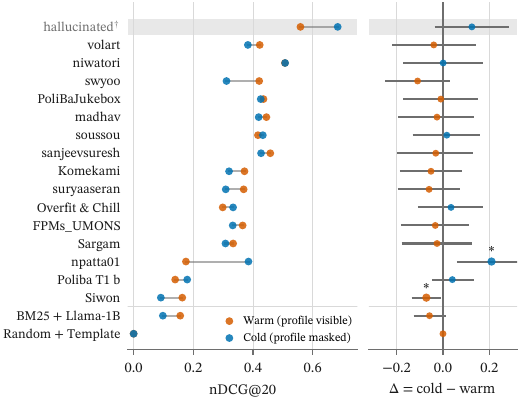}
\vspace{-5mm}\caption{Warm versus cold nDCG@20 per system (left) and the cold-minus-warm contrast with stratified-bootstrap 95\% intervals (right). $^{*}$~interval excludes zero. $\dagger$ The evaluation-data-exposed submission.}
\label{fig:warmcold}
\end{figure}

\begin{figure*}[t]
\centering
\includegraphics[width=\textwidth]{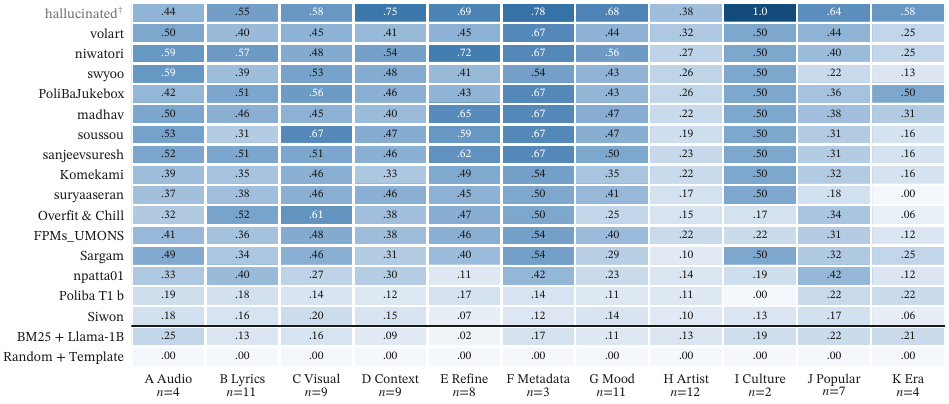}
\vspace{-5mm}
\caption{Turn-macro nDCG@20 by conversation-goal topic. $\dagger$ The evaluation-data-exposed submission.}
\label{fig:goaltopic}
\end{figure*}

\begin{figure}[t]
\centering
\includegraphics[width=\columnwidth]{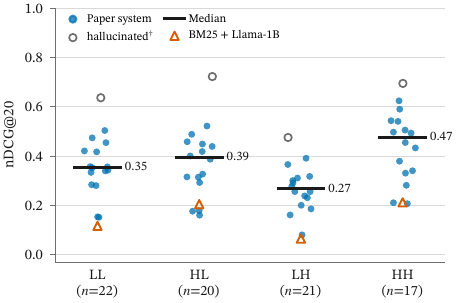}
\vspace{-5mm}
\caption{Per-system nDCG@20 by goal-specificity regime. Precise lookups (HH) are easiest and vague single-target requests (LH) hardest. $\dagger$ The evaluation-data-exposed submission.}
\label{fig:specificity}
\end{figure}

\subsection{Warm Start versus Cold Start}

Figure~\ref{fig:warmcold} compares nDCG@20 across 40 warm sessions with visible user IDs and profiles and 40 cold sessions in which both are masked. It reveals three regimes. First, hallucinated's cold score (0.684) exceeds its warm score (0.559). Before its disclosure this would read as remarkable cold-start robustness from query/session evidence, worst-slice CVaR tuning, and continuity heuristics~\cite{alkhetiar2026talk}. Given that its candidate fusion and reranker were tuned and trained on Blind-B conversation turns, the asymmetry cannot be attributed to architecture alone.

Second, the only non-exposed contrast whose interval excludes zero on the cold side belongs to npatta01 ($+0.210$): its typed-state, text-first retrieval treats the profile as one optional branch~\cite{pattaniyil2026state}, although the asymmetry equally reflects a weak warm-side score (0.175), so it is evidence of profile independence rather than of overall strength. More broadly, niwatori is essentially invariant (0.507 versus 0.508), while soussou and Overfit \& Chill are modestly higher on cold. Their shared property is that identity is not foundational: niwatori and soussou use no profile or user-identity signal at all, drawing history only from tracks observed in the dialogue~\cite{wakatsuki2026practical,soussou2026multisource}, and Overfit \& Chill's user-to-item BPR channel is one of ten in its union~\cite{sundrani2026minimaestro}.

Third, most systems favor warm sessions, but only Siwon's $-0.072$ contrast excludes zero in that direction~\cite{lee2026empirical}. Among point estimates, swyoo shows the largest warm advantage ($-0.110$), whereas PoliBaJukebox is nearly invariant despite using no profile features and reweighting training toward blind-like sessions~\cite{lops2026distribution}. The profile-agnostic BM25 baseline shows a similar warm advantage ($-0.058$; 0.098 cold versus 0.156 warm), suggesting that the masked half is intrinsically harder. Thus, the dominant pattern is a cold-session penalty, although it cannot be attributed solely to profile removal. Wide intervals across the 40-session halves preclude a definitive cold-start ranking.

\subsection{Goal Specificity}

Across the 16 systems, median nDCG is 0.355 on LL, 0.394 on HL, 0.267 on LH, and 0.474 on HH (Figure~\ref{fig:specificity}). HH benefits from explicit titles, artists, albums, and other catalog-resolvable constraints; nearly every pipeline contains exact matching or entity expansion. In contrast, LH asks the system to recover one or a few intended tracks from vague clues. This combines low observable information with a single-target metric, matching the information-limit diagnosis that dialogue identifies a neighborhood more reliably than the logged point within it~\cite{volgin2026frozen}. The same LH dip appears in the BM25 baseline (0.066 versus 0.213 on HH), so the regime ordering is a property of the task, not of any modeling family.

The gap also explains why an exact-target module is best treated as a routed specialist rather than a universal semantic reranker. High-precision catalog resolution can help HH~\cite{rai2026dialogue,suresh2026label}, while LH needs diversified recall, uncertainty-aware ranking, or clarification. The exposed hallucinated submission posts the highest value on all four slices; among non-exposed systems, niwatori leads LL, LH, and HH while volart leads HL. Thus no one modality explains performance across the specificity axis.

\subsection{Goal Topic}

Figure~\ref{fig:goaltopic} summarizes topic difficulty across every submission. Artist/\allowbreak discography (H, median 0.219) and temporal/era (K, 0.158) are hardest, followed by social/popularity context (J, 0.319). The artist result is not evidence that entity matching fails; rather, many H goals ask for discovery within or beyond a discography, where the logged continuation policy, artist repetition, and user request may conflict~\cite{patil2026routing,terai2026auditing}. Era queries combine fuzzy temporal language with incomplete or noisy release metadata.

Visual (C, median 0.472) and metadata-rich (F, 0.544) appear easier, with soussou strongest on C, consistent with its explicit last-item and history-image retrievers~\cite{soussou2026multisource}. These slices contain nine and three sessions, respectively, so this is a hypothesis for a larger test, not an ablation. The same caveat applies to culture/geography (I): the exposed hallucinated submission scores a perfect 1.0, but $n=2$. Audio (A) is led jointly by niwatori and swyoo (0.589 each), and lyrics (B) by niwatori, whose broad source union covers both lexical and semantic evidence~\cite{wakatsuki2026practical}.

\section{Reusable Design Principles}
\label{sec:principles}
The system papers and our stratified analysis yield four evidence-based design principles for future conversational music recommenders. We ground each principle in reported system components or controlled ablations, distinguish submitted methods from analysis-only findings, and use the organizer-side slices in Section~\ref{sec:protocol} as supporting rather than causal evidence.

\vspace{2mm}\noindent\textbf{1. Combine candidates from multiple retrievers, then learn which sources to trust.}
Use lexical, dense, entity, history, co-occurrence, and optional identity or multimodal sources as parallel candidate generators: each fails in a different regime, so their union sets a high recall ceiling cheaply. The reranker should then use source-specific evidence: which retrievers proposed a track, at what rank, with what score, and how many retrievers agree. Ten of the 16 accepted systems pass such signals to a tree-based learning-to-rank model~\cite{wakatsuki2026practical,volgin2026frozen,lops2026distribution,yoo2026twoviews,seran2026dialogue,terai2026auditing,soussou2026multisource,sundrani2026minimaestro,pattaniyil2026state,alkhetiar2026talk}. niwatori provides a controlled comparison over the same candidate union: RRF reaches 0.1715 nDCG@20, a LambdaRank model using only context--track features reaches 0.1832, and adding per-retriever rank, score, presence, and agreement features raises it to 0.1994~\cite{wakatsuki2026practical}. When no training budget exists, weighted RRF over the same union remains a strong fallback and candidate-pool constructor~\cite{rai2026dialogue,lee2026empirical}.

\vspace{2mm}\noindent\textbf{2. Use conversation and item signals to handle cold-start.}
Cold-start robustness comes from grounding the ranking pipeline in information that remains observable without a user profile. niwatori does not use user IDs, profiles, or user embeddings for ranking. Its BM25, TF--IDF, tag/entity, and two-tower sources retrieve from the current message and observed dialogue, while artist/album expansion and co-occurrence/transition sources use tracks already observed within the session~\cite{wakatsuki2026practical}. npatta01 illustrates a similar fallback architecture: it compiles the dialogue into a typed conversation state and makes user-CF only one of eleven retrieval branches, so text, entity, and item-based branches can operate without it~\cite{pattaniyil2026state}. This is architectural rather than performance evidence, since its paper reports that an in-sample estimate of 0.3844 nDCG@20 fell to 0.2032 under out-of-fold evaluation. Stronger validation comes from volart, which masks the goal and half of the profiles on held-out data and finds essentially no loss after the corresponding features are nulled (goal $\Delta=-0.0025$, $p=0.31$)~\cite{volgin2026frozen}. The converse case is Overfit \& Chill: two unavailable goal fields collapsed to an unseen sentinel value on Blind-B, and retraining without them raised the composite from 0.42 to 0.48~\cite{sundrani2026minimaestro}. Together, these results favor conversation- and item-based retrieval as the default path, with identity, profiles, and collaborative signals treated as optional additions.

\vspace{2mm}\noindent\textbf{3. Use targeted intent detectors for specific ranking actions.}
The organizer-side goal categories and HH/HL/LH/LL specificity labels used in our analysis were not visible in Blind-B, so the evidence supports detecting narrow, actionable intents rather than recovering the full hidden taxonomy. sanjeevsuresh detects exact-song and version requests with regular expressions over quoted titles and imperative phrases, resolves them against title and artist fields in the catalog, and abstains when the match is ambiguous. The detector gates a specialist reranker trained with request-satisfying targets in addition to the official labels, raising request-satisfying nDCG@20 from 0.523 to 0.802 on the 43-case development exact/version conflict slice, while preserving official whole-split nDCG@20 (0.1908 → 0.1914)~\cite{suresh2026label}. For continuation versus discovery, Compass trains a logistic classifier on the dialogue-query embedding to predict whether the target artist already appears in the session history. Its predicted probability scales the artist/album continuity boost, reaching AUC 0.788 and improving the novel-artist slice in local analysis; however, this gate was analysis-only and did not yield an overall Blind-B gain~\cite{patil2026routing}. Komekami likewise uses high-precision lexical rules for phrases such as ``a different artist'' or ``something new,'' but only to audit artist-change intent rather than route its submitted ranker~\cite{terai2026auditing}. These results favor high-precision detectors that abstain under ambiguity and trigger a defined ranking action; they do not establish general goal or specificity prediction.

\vspace{2mm}\noindent\textbf{4. Do not focus only on the current query; consider the multi-turn context.}
Recommendation intent is distributed across the dialogue: the current request may refine, reject, or refer back to preferences and tracks introduced in earlier turns. Systems should therefore model the conversation as a whole rather than reduce it to the latest utterance. In Siwon's controlled dense-retrieval comparison, the full-dialogue query reaches 0.091 nDCG@20, while three request-focused variants score 0.089, 0.085, and 0.082; full-dialogue BM25 is stronger still at 0.118~\cite{lee2026empirical}. swyoo's fusion of a conversational pool with its hybrid pool raises nDCG@20 from 0.2165 to 0.2198~\cite{yoo2026twoviews}. soussou provides a complementary ablation: removing its contextual dense retriever lowers nDCG@20 from 0.1851 to 0.1778, while removing its recent-utterance dense retriever lowers it to 0.1825~\cite{soussou2026multisource}. Our turn-depth analysis is consistent with this principle because later turns remain harder, but it does not by itself identify inadequate context modeling as the cause.

\section{Limitations and Future Challenges}
\label{sec:limitations}

\vspace{2mm}\noindent\textbf{Evaluation-data exposure.}
The winning submission's paper discloses that its RRF fusion weights were tuned to maximize recall on held-out Blind-B target turns and that its XGBoost reranker was model-selected on the last available turn of each Blind-B session and, for the final submission, trained on ``all available Blind-B turns except those used for validation''~\cite{alkhetiar2026talk}. Although the hidden target tracks themselves were never revealed, the conversational history of every evaluation session, including the sessions whose profiles were masked, entered training. This is outside the intended scenario, in which Blind-B measures generalization to unseen conversations, and it plausibly inflates both the overall score and the cold-start result in Section~\ref{sec:protocol}. We keep the official ranking intact, because the challenge rules did not explicitly forbid transductive use of the released blind inputs, but we gray out the affected rows in every table and figure and exclude the system when naming slice leaders. Future editions should state an explicit data-usage contract (no training, tuning, or model selection on blind-phase inputs), require a data-usage declaration at submission time, and audit winning pipelines against it before finalizing awards.

\vspace{2mm}\noindent\textbf{Dataset integrity: Reverse-engineering risk.}
Because the challenge dataset (TalkPlayData-Challenge) is synthetically generated from LFM-2b listening histories, the benchmark may retain track sequence correspondences to its source data. During the challenge, we received a report that similarities between multi-turn conversational patterns and listening histories could be exploited to partially reconstruct the latent recommendation pool. Although this does not directly reveal the hidden target, it can reduce the effective search space and highlights the importance of dataset integrity in challenge benchmarks, particularly as synthetic datasets derived from existing corpora become more common.

To address this concern, the organizers conducted additional verification of the award-winning submissions, including not only inference reproduction but also independent reconstruction of the submitted systems from their code and configurations. This process provided assurance that the reported performance could be reproduced without relying on reverse engineering of the challenge dataset. Future editions should explicitly audit source-data linkages and reconstructability as part of benchmark design and validation. We thank the members of team hallucinated~\cite{alkhetiar2026talk} for identifying and reporting this issue.

\vspace{2mm}\noindent\textbf{Dataset limitation: Single-ground-truth scenario.}
Each turn has one target selected by the synthetic recommender from a latent session pool, so open-ended requests may admit many equally useful tracks that the metric counts as failures, and exact requests can conflict with the logged choice (Section~\ref{sec:principles}); npatta01 further reports that two LLM judges relabeling training turns disagree roughly 20\% of the time~\cite{pattaniyil2026state}. The one-label problem is most acute in the LH regime, the hardest slice for nearly every system including the BM25 baseline: when a vague query is paired with a single logged target, the metric cannot separate a wrong recommendation from a valid alternative that the policy did not log. In such cases a real assistant would likely ask a clarifying question rather than commit to a list. Future releases should combine the logged action with catalog-resolved constraints and graded, multi-label relevance, report target recovery and request satisfaction separately, and permit an explicit clarification action scored by the information it recovers.

\vspace{2mm}\noindent\textbf{Evaluation limitation: Teacher-forced, synthetic multi-turn evaluation.}
Every target turn is evaluated against the logged history: systems condition on the synthetic policy's past recommendations, never on their own, so error compounding, recovery from a bad suggestion, and long-horizon preference tracking are untested. The benchmark ultimately measures recovery of a synthetic policy's next track rather than the long-term utility of an interactive recommender. A next edition should complement teacher-forced snapshots with a reproducible simulator or a controlled user study in which systems may ask clarifying questions and receive feedback, scored by turns-to-success, constraint satisfaction, and user effort.

\subsection{Encountered Issues}
\label{sec:issue}

The main issue we faced concerned the scope of the recommendation catalog. The metadata released to participants contained two splits, \texttt{all\_tracks} and \texttt{test\_tracks}, with the second one originally included to experiment with the appropriate scope of tracks for the evaluation. Submissions were expected to rank tracks from the full \texttt{all\_tracks} catalog. This requirement was stated in the baseline GitHub repository from the beginning of the challenge. It was not added to the Codabench description until the start of Blind-B, resulting in inconsistent guidance across the two official sources.%. As a result, the two official sources gave different instructions for part of the competition.

Restricting the recommendations to \texttt{test\_tracks} drops the candidate pool to a subset much closer to the evaluation data, which inflates the ranking performance. The inconsistency was identified near the end of the competition after being reported by a participating team. Teams affected by the issue were allowed to resubmit using the full catalog, and the \texttt{test\_tracks} split was removed from the Hugging Face repository to prevent further confusion. All results reported in Table~\ref{tab:leaderboard} correspond to submissions evaluated over the complete \texttt{all\_tracks} catalog.

\section{Conclusion}

The RecSys Challenge 2026 established a shared benchmark for conversational music recommendation, bringing together conversational understanding, large-catalog retrieval, learned ranking, and grounded response generation within a single evaluation setting. Across 16 accepted systems, the clearest retrieval pattern is to combine complementary candidate sources and preserve source-specific ranks, scores, and presence together with cross-source agreement for learned reranking. For cold-start users, the more robust default is to retrieve from the conversation and observable item evidence while treating identity, profiles, and collaborative signals as optional additions.

Our organizer-side analysis also shows that one recommendation policy does not fit every request. The available evidence favors narrow, high-precision intent detectors that abstain under ambiguity and trigger a defined ranking action, rather than attempting to reconstruct a complete hidden goal taxonomy. It also favors modeling the full multi-turn dialogue: current requests often refine, reject, or refer back to evidence from earlier turns, and later turns remain difficult even for strong systems. Vague single-target requests and artist or era discovery remain especially challenging, underscoring the need to represent uncertainty rather than force every conversation into one confident target.

Finally, the challenge exposed important limitations of the benchmark itself, including evaluation-data exposure, single-ground-truth relevance, and teacher-forced synthetic multi-turn evaluation. Future benchmarks should adopt clearer data-usage protocols, a single authoritative task specification kept in sync across all channels, richer and potentially graded relevance judgments, stronger dataset-integrity checks, and genuinely interactive evaluation. Moving beyond recovery of a logged target toward measuring request satisfaction and conversational utility will be essential for evaluating whether future systems provide recommendations that are both relevant and useful.

% alredy thanks at, Dataset integrity part

% The authors extend their gratitude to the challenge participants and the RecSys organizers for their participation and support. The authors also thank the members of the team hallucinated~\cite{alkhetiar2026talk} who identified the data leak issue. We also thank the sponsoring organizations of the RecSys Challenge 2026.

\begin{acks}
We thank everyone who contributed to organizing the RecSys Challenge 2026, as well as all participating teams for their engagement and contributions. We are also deeply grateful to Sony Group Corporation, SiriusXM, and Deezer for their generous support of the challenge prizes, which helped recognize the outstanding efforts of the participating teams.
\end{acks}

\bibliographystyle{ACM-Reference-Format}
\bibliography{references}

\end{document}